\documentclass[aps,pra,epsfigure,twocolumn]{revtex4}
\usepackage{dcolumn}    
\usepackage{bm} 
\usepackage{graphicx}
\usepackage{amsmath}    
\usepackage{latexsym}
\usepackage{amsfonts}   
\usepackage{amssymb}
\usepackage{array}      
\usepackage{epsfig}
\usepackage{txfonts}
\usepackage{color}
\usepackage[colorlinks=true,linkcolor=blue,urlcolor=blue,citecolor=blue]{hyperref}
\usepackage{hyperref}

\newcommand{\ket}[1]{\left\vert#1\right\rangle}
\newcommand{\bra}[1]{\left\langle#1\right\vert}

\begin{document}
 \title{Criticality revealed through quench dynamics in the Lipkin-Meshkov-Glick model}
       
\author{Steve Campbell}
\affiliation{Centre for Theoretical Atomic, Molecular and Optical Physics, Queen's University Belfast, Belfast BT7 1NN, United Kingdom}

\begin{abstract}
We examine the dynamics after a sudden quench in the magnetic field of the Lipkin-Meshkov-Glick model. Starting from the groundstate and by employing the time-dependent fidelity, we see manifestly different dynamics are present if the system is quenched through the critical point. Furthermore, we show that the average work shows no sensitivity to the quantum phase transition, however the free energy and irreversible work show markedly different rates of change in each phase. Finally, we assess the spectral function showing the fundamental excitations that dictate the dynamics of the post-quenched system, further highlighting the qualitative differences between the dynamics in the two phases.
\end{abstract}
\date{\today}
\maketitle

\section{Introduction}
Phase transitions are an interesting trait of physical systems. In particular, when dealing with quantum many-body set-ups it is curious that a well defined microscopic description can lead to non-trivial and singular behaviours in the thermodynamic limit. The study of the equilibrium properties of such systems is well established in, for example, exactly solvable one-dimensional spin chains~\cite{Sachdev}. With the help of several tools from quantum information, notably quantum correlations, the implications of the presence of quantum phase transitions (QPTs) have been explored~\cite{CorrelationsQPT}.

Beyond the static properties, there is a growing interest in studying the dynamics of many-body quantum systems. This is further catalysed by the increasing interest in understanding the thermodynamic properties of genuinely quantum systems~\cite{GooldJPA}. Despite the significantly more involved nature of studying dynamics, remarkable progress has been made in elucidating the behaviour of important quantities, e.g. (irreversible) work, entropy production, and residual energy, when a many-body system is evolved through its critical point. The most frequent evolution considered is that of a ``sudden quench" of the order parameter~\cite{EchoQuenches,SilvaPRL,DornerPRL,LorenzoPRX,GabrielePRB,DasPRB} (however finite time protocols have also been addressed~\cite{SaroPRB}). Focusing on such a sudden change allows us to capture the salient features of the ensuing non-equilibrium dynamics, while leaving the study of more qualitative differences to a more involved temporal analysis. Typically, the sudden change to the Hamiltonian kicks the system out of equilibrium and can lead to interesting consequences. Notably, for the Ising model the dynamics of a sudden quench have been explored and it has been shown that the irreversible entropy production provides signals of the presence of the equilibrium QPT~\cite{SilvaPRL}, and also can be used to explain emergent phenomena such as the vanishing gap between ground and first excited energy levels in the thermodynamic limit~\cite{DornerPRL}. Recently, the irreversible work was shown to faithfully capture the critical features even for so-called impurity QPTs~\cite{GabrielePRB}. 

In this work we add to this endeavour by studying the Lipkin-Meshkiv-Glick model~\cite{LMG65}. The model has attracted substantial interest as it serves as the paradigmatic example of an infinite range interacting system. It can be solved in the thermodynamic limit and exhibits a complex phase diagram~\cite{VidalPRLPRE,CastanosPRB}. We will be interested in exploring how clear signatures of the equilibrium QPT is manifest in the dynamics when the model is quenched through its critical point. We remark that the evolution of this model through its QPT have been studied previously in Ref.~\cite{DasPRB} wherein the equal time-order parameter correlation function was examined and Ref.~\cite{SaroPRB} where the adiabatic dynamics were explored. Our study is set apart from these as it seeks to establish a rigorous link between the thermodynamic quantities such as work and free energy, with the presence of the known QPT. By exploiting the time-dependent fidelity we show that the dynamics are manifestly different when the quench is restricted to a particular phase compared to when the system is quenched through the critical point. More interestingly, we explicitly show that while the average work performed on the system due to the quench is blind to the QPT, the free energy, and therefore the irreversible work, appears acutely sensitive, showing a markedly different rates of change. Finally, we use the spectral function to further understand the fundamental excitations governing the dynamics of the system.

The remainder of the paper is organised as follows. In Sec.~\ref{model} we present the model and introduce the quantities that will be of interest to our analysis. In Sec.~\ref{results} we explore these quantities for quenching the system across the critical point and show that the time-dependent fidelity exhibits interesting features that reveal the critical nature of the system as well as showing that the irreversible work neatly reveals the QPT. Sec.~\ref{spectral} assess the spectral function for various quenches. Finally, Sec.~\ref{conclusions} we present our conclusions and some discussions on our results.

\section{The Model and Figures of Merit}
\label{model}
\begin{figure}[t]
\includegraphics[width=0.85\columnwidth]{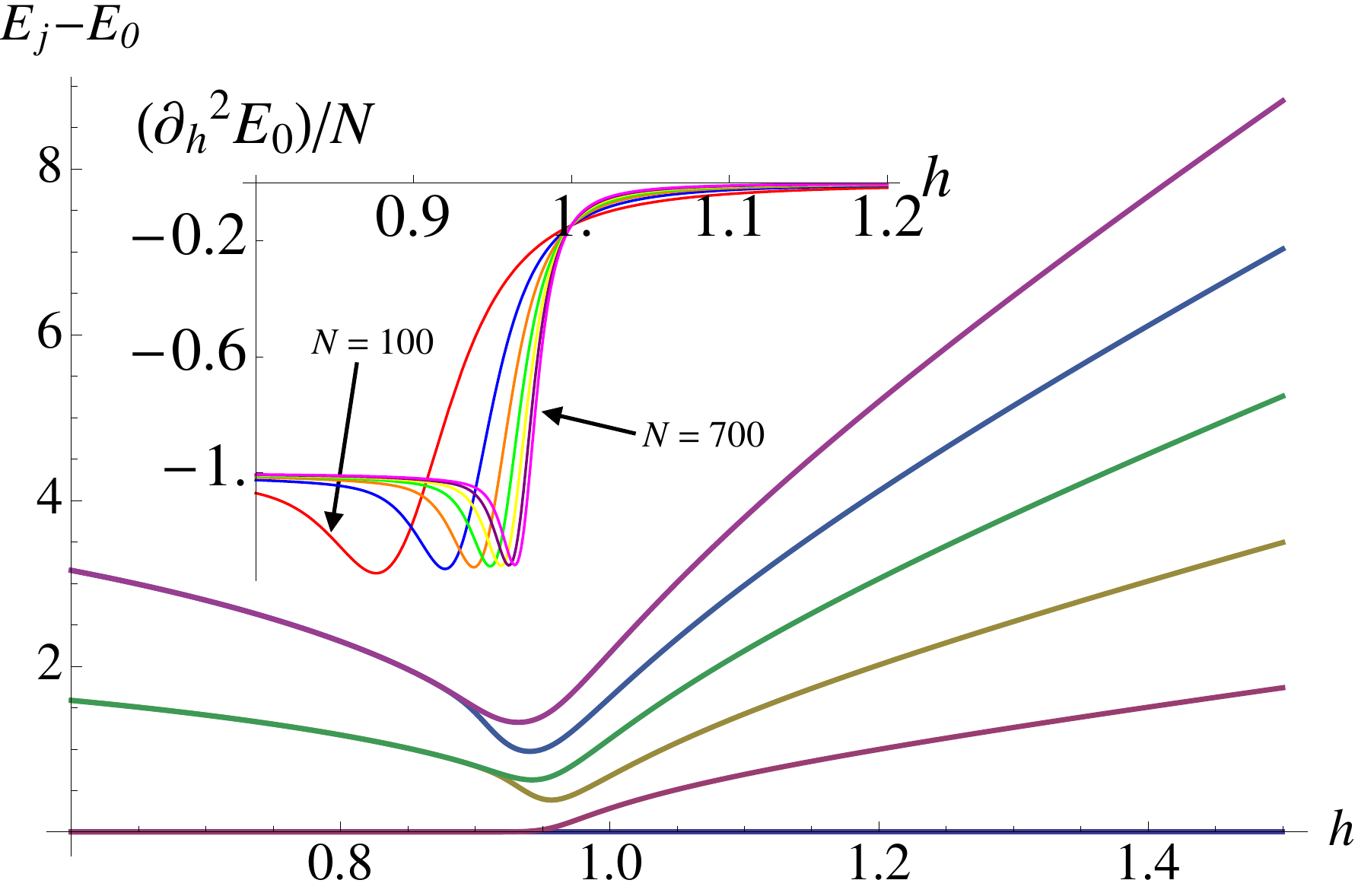}
\caption{(Color online) Energy difference between the ground state and the first 5 excited states for $N=400$ plotted against magnetic field strength $h$. {\it Inset:} Second derivative of the ground state energy per site for $N=100$ (red) to $700$ (magenta) in steps of 100.}
\label{fig1}
\end{figure}

We consider the ferromagnetic spin-1/2 Lipkin-Meshkov-Glick (LMG) model in a transverse field,
\begin{equation}
\mathcal{H} = -\frac{1}{N}\left( \sum_{i<j} \sigma_x^i \otimes \sigma_x^j+\gamma \sigma_y^i \otimes \sigma_y^j \right) - 
h \sum_{i} \sigma_z^i,
\end{equation}
with $\sigma_{x,y,z}$ the Pauli spin-operators, $h$ the magnetic field strength, and $\gamma$ the anisotropy parameter (which we set to zero for simplicity in our simulations, however we remark that qualitatively similar results can be obtained for any $0\leq\gamma<1$, cfr. the appendix). By considering the collective spin operators $S_\alpha=\sum_i \sigma_{\alpha}^i/2$ with $\alpha=\{x,y,z\}$, up to a constant energy shift, the model can be written as
\begin{equation}
\label{collspinLMG}
\mathcal{H} =-\frac{2}{N}\left( S_x^2 + \gamma S_y^2 \right) - 2h S_z.
\end{equation}
In the following we work in the basis of maximum angular momentum (which is a constant of motion) and using the eigenstates of $S_z$ we can diagonalize Eq.~\eqref{collspinLMG} to find the complete spectrum (see e.g. Refs~\cite{VidalPRLPRE,CastanosPRB} for further details). In Fig.~\ref{fig1} we show the energy difference between the 5 lowest excited states and the ground state against $h$. We see when $h>1$ each energy level is distinct. As $h$ is decreased the gap between the first excited state and the ground state closes, and similarly the energy gap between subsequent pairs of excited states also closes. However, an important remark, only when $N\to \infty$ does the gap vanish and all eigenstates become doubly degenerate. Hence, for any finite size there is a small difference between the ground and first excited states~\cite{SaroPRB}.

The inset in Fig.~\ref{fig1} shows the second derivative with respect to $h$ of the ground state energy (per site) for system sizes ranging from $N=100$ to $N=700$. We see the emergence of a discontinuity appearing, thus signalling the known second order QPT at $h=1$~\cite{VidalPRLPRE}. We are interested in studying the dynamics when the ground state of one phase is evolved using the propagator of another. In what follows we will assume the system is initialized in the ground state of Eq.~\eqref{collspinLMG} corresponding to $h=h_i$. At time $t=0$, we quench the field strength $h_i \to h_f$ and we evolve the initial state according to the new Hamiltonian $\mathcal{H}_f$, so that
\begin{equation}
\ket{\psi(t)} = e^{-i \mathcal{H}_f t} \ket{\psi(0)}.
\end{equation}
Using this we can readily evaluate the time dependent overlap
\begin{equation}
\mathcal{O} = \bra{\psi(0)} \psi(t) \big>.
\end{equation}
This quantity will be central to our analysis as it allows us to access several important quantities that indicate that signatures of the equilibrium QPT are clearly manifest in the system's evolution. 

A particularly important quantity will be the time-dependent fidelity (TDF)
\begin{equation}
\mathcal{L} = \vert \mathcal{O} \vert^2,
\end{equation}
which quantifies how different the evolved state is compared to the initial one (we remark this quantity is sometimes referred to as the Loschmidt echo in the literature). The TDF has already proven to be a useful tool in studying critical dynamics~\cite{EchoQuenches}. Additionally, we can determine the average work due to the sudden quench~\cite{LorenzoPRX,CampbellMarchFogarty}
\begin{equation}
\big< W \big> =  \sum_j (E_j^f -E_0^i) \big\vert \big<\psi_0^i \vert\psi_j^f\big> \big\vert^2
\end{equation}
where $E_j^f$ and $\vert\psi_j^f \big>$ are the $j$-th eigenenergy and eigenstate of the post-quench Hamiltonian, and $E_0^i$ and $\vert \psi_0^i\big>$ are ground state energy and ground state for the initial Hamiltonian. The sudden nature of the quench drives the system out of equilibrium, and thus introduces a degree of irreversibility of the process. We can quantitatively define the irreversible work as~\cite{GabrielePRB,DeffnerPRL,CampbellMarchFogarty}
\begin{equation}
\big< W_\text{irr} \big> = \big< W \big> - \Delta F  
\end{equation}
where $\Delta F$ is the free energy difference. By considering closed dynamics, and since we assume our system begins in the ground state of the initial Hamiltonian, $\Delta F$ is simply given by the difference between the post- and pre-quench ground state energies, i.e. $\Delta F = E_0^f - E_0^i$. 

\section{Quench Dynamics Across a Quantum Critical Point}
\label{results}
\subsection{From the Paramagnetic to the Ferromagnetic Phase}
\begin{figure}[t]
\includegraphics[width=0.85\columnwidth]{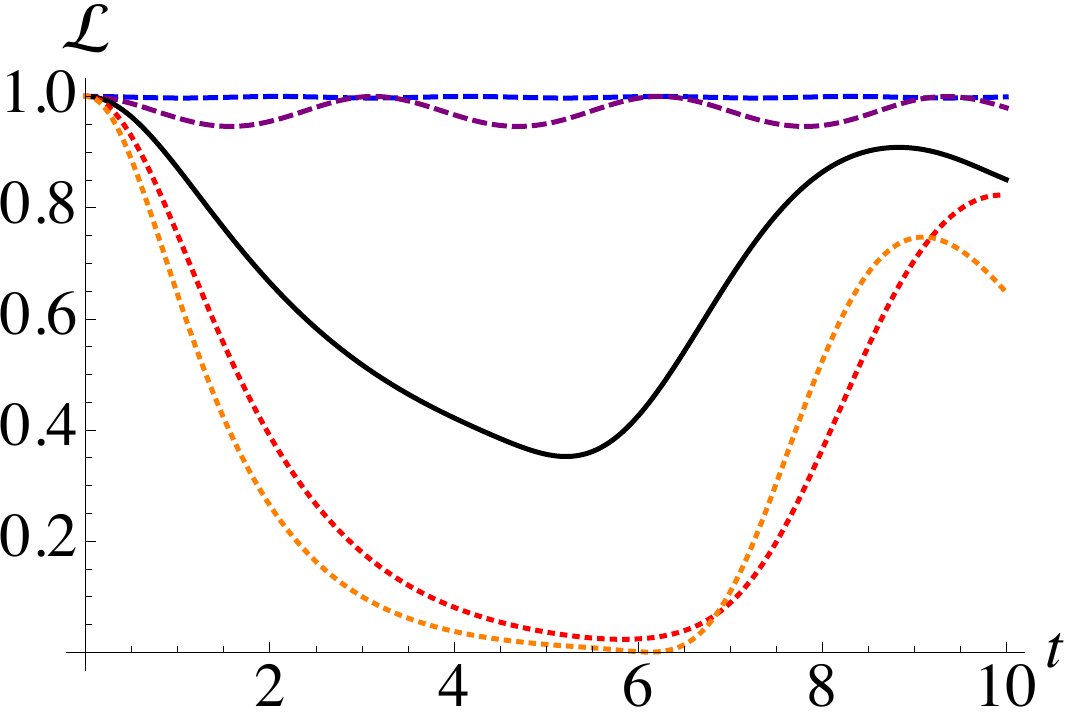}
\caption{(Color online) Time-dependent fidelity for a quench from $h_i=1.5$ to $h_f \in [0.6,1.4]$ with $N=400$. The dashed curves are for $h_f>1$ with $h_f=1.4$ (topmost, blue) and $h_f=1.2$ (purple). The lowest two dotted curves are for $h_f<1$ with $h_f=0.8$ (red) and $h_f=0.6$ (bottom-most orange). The solid black curve is for $h_f=1$.}
\label{fig2}
\end{figure}

We begin analysing the case of quenching from the paramagnetic phase, setting $h_i=1.5$. In this regime there is a significant energy difference between the ground state and the first excited state, cfr Fig.~\ref{fig1}. In Fig.~\ref{fig2} we show the TDF for $N=400$ and the thick black curve corresponds to $h_f=1$. The dashed curves above this are for quenches that evolve the initial state taking a value of $h_f$ that is still in the paramagnetic phase. We clearly see the regular oscillatory behaviour persists even when quenching close to the critical point. When we evolve the state using $h_f<1$ (lower dotted curves) we see the dynamics loses the clean periodic behaviour, and dynamically the TDF no longer reaches unity. Additionally there is a significant decrease in the values of $\mathcal{L}$, even in some cases reaching exactly 0 indicating that the evolved state is orthogonal to the initial state.

\begin{figure}[t]
\includegraphics[width=0.95\columnwidth]{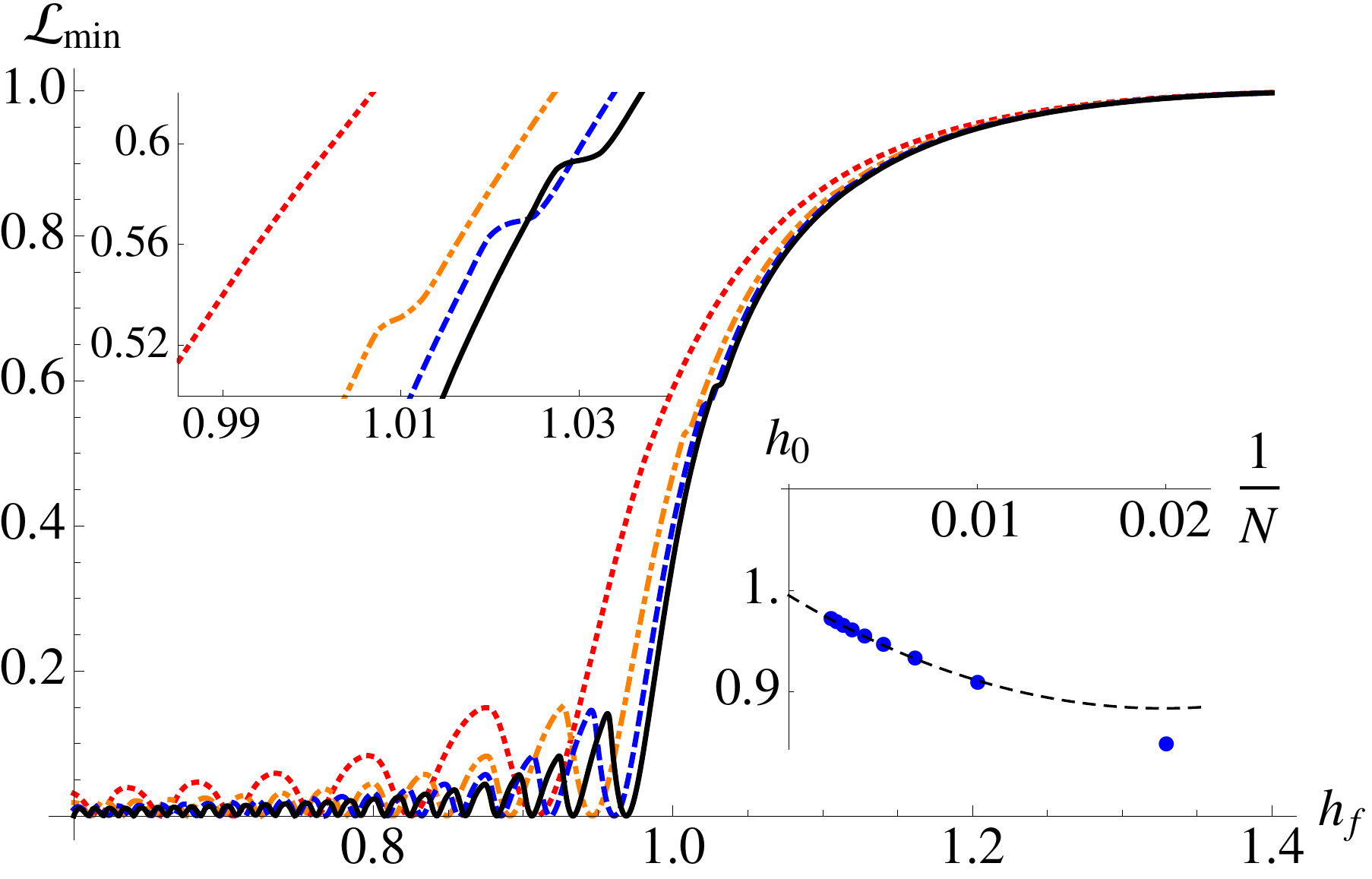}
\caption{(Color online) {\it Main Panel:} minimum value of TDF achieved for $t\in(0,10)$ quenching the field from $h_i=1.5$ to $h_f$. Each curve from left to right corresponds to an increasing size of $N=100$ (dotted, red), 200 (dash-dotted, orange), 300 (dashed, blue) and 400 (solid, black). {\it Upper left inset:} Zoomed in cross-section of main panel. {\it Lower right inset:} Finite size scaling for the first value of field at which the $\mathcal{L}_\text{min}=0$.}
\label{fig3}
\end{figure}

The precise value at which the TDF reaches zero is examined in Fig.~\ref{fig3}. We determine the minimum value of TDF observed, $\mathcal{L}_\text{min}$, within the same time window for Fig.~\ref{fig2} against $h_f$. Clearly, when the quench is small, e.g. $h_f\in(1.2,1.4)$ the minimum value of TDF is still quite large. As the strength of the quench is increased we find this minimum value decreases. Interestingly, $\mathcal{L}_\text{min}=0$ only when $h_f<1$. We see an oscillatory behaviour appearing, however the amplitude of the oscillations is decreasing as the system size is increased. Furthermore, the first value of $h_f$ where $\mathcal{L}_\text{min}=0$, denoted $h_0$, shifts closer to 1 as we increase the system size. Through a finite size scaling with a quadratic fit, the lower right inset shows that this accurately determines the critical point. Such a result is remarkable as it clearly indicates that the equilibrium QPT can be witnessed by the occurrence of dynamical orthogonality. 

A final peculiarity appears in studying $\mathcal{L}_\text{min}$, for $N>200$ a kink appears close to the critical point as shown in the upper left inset of Fig.~\ref{fig3}. However, as the system size is increased, this feature appears to move further from the critical value. 

\begin{figure}[t]
{\bf (a)}\\
\includegraphics[width=0.8\columnwidth]{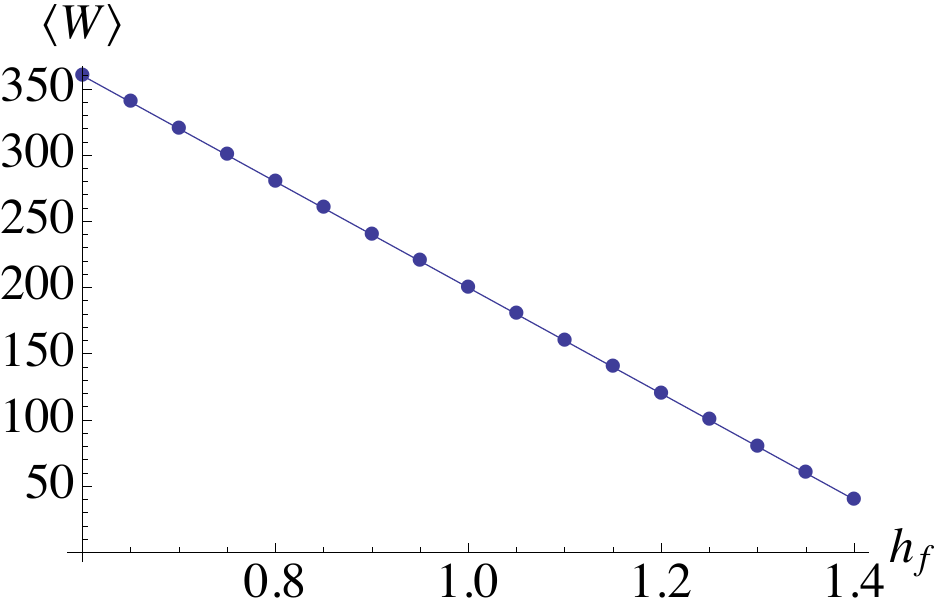} \\
{\bf (b)}\\
\includegraphics[width=0.8\columnwidth]{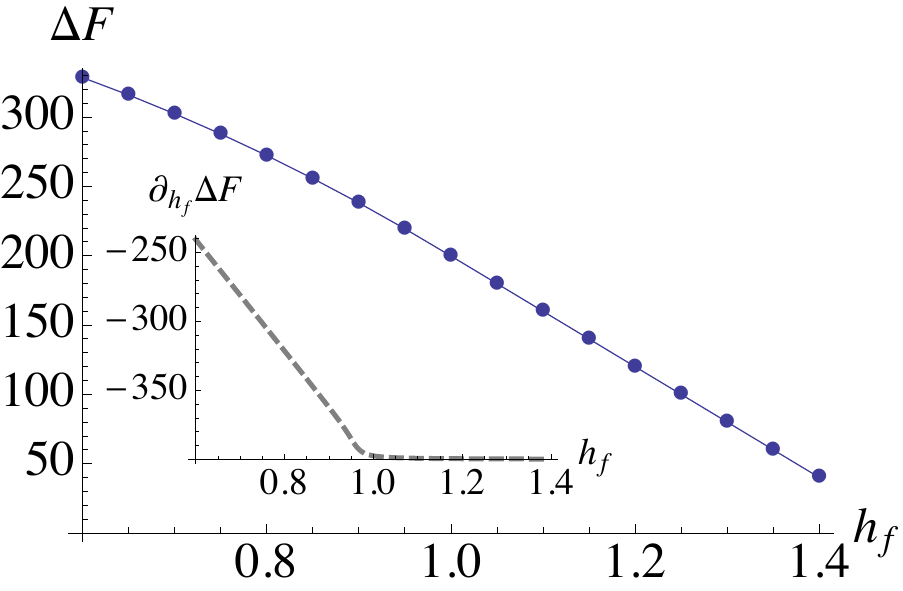}\\
{\bf (c)} \\
\includegraphics[width=0.8\columnwidth]{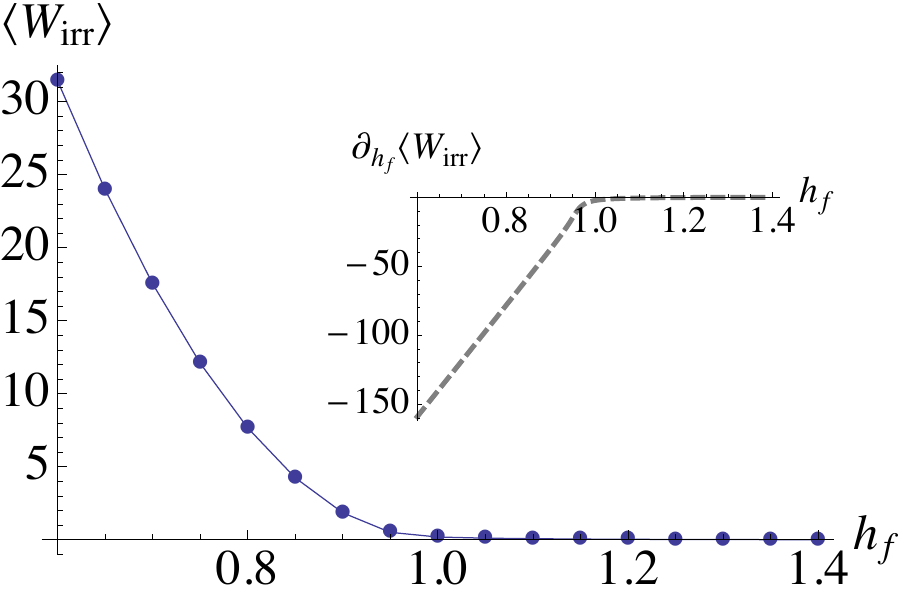}
\caption{(Color online) {\bf (a)} Average Work {\bf (b)} Free energy {\bf (c)} Irreversible work for a quench from $h_i=1.5$ to $h_f$. The insets of panels {\bf (b)} and {\bf (c)} are the first derivative of the functions. In all panels $N=400$.}
\label{fig4}
\end{figure}

From Fig.~\ref{fig2} it is clear that for small quenches the system dynamically comes close to the initial state, however for larger quenches this is no longer the case, indicating that a degree of irreversibility has been introduced into the system~\cite{BatalhaoPRL}. In Fig.~\ref{fig4} {\bf (a)} we examine the average work done,  $\big< W \big>$, against $h_f$. We find $\big< W \big>$ is linearly dependent on the value of $h_f$, the larger the quench $h_i \to h_f$, the more work is done. In panels {\bf (b)} and {\bf (c)} we show the free energy and average irreversible work, respectively. Again, the free energy increases as we increase the size of the quench. However in the inset we examine its rate of change, and we see for $h_f>1$ this rate is linear, and there is a sudden change near $h_f\sim1$. While it is not surprising that the free energy exhibits a non-trivial behavior as we go through the critical point in light of the fact that is defined in terms of the ground state energy, the fact that both $\big< W \big>$ and $\Delta F$ are of the same order of magnitude leads to a trade off between the two quantities. This has interesting consequences for the irreversibility of the process captured by $\big<W_\text{irr} \big>$. Panel {\bf (c)} shows that when the quench is small and confined to the same phase as the initial state, $\big< W_\text{irr} \big> = 0$, indicating that the process is fully reversible, as confirmed by the behavior of the TDF which achieves values of unity during the dynamics. For large quenches, when the system is evolved according to a Hamiltonian in the ferromagnetic phase the average irreversible work becomes non-zero, and the degree of irreversibility grows as the magnitude of the quench increases.

\begin{figure}[t]
\includegraphics[width=0.8\columnwidth]{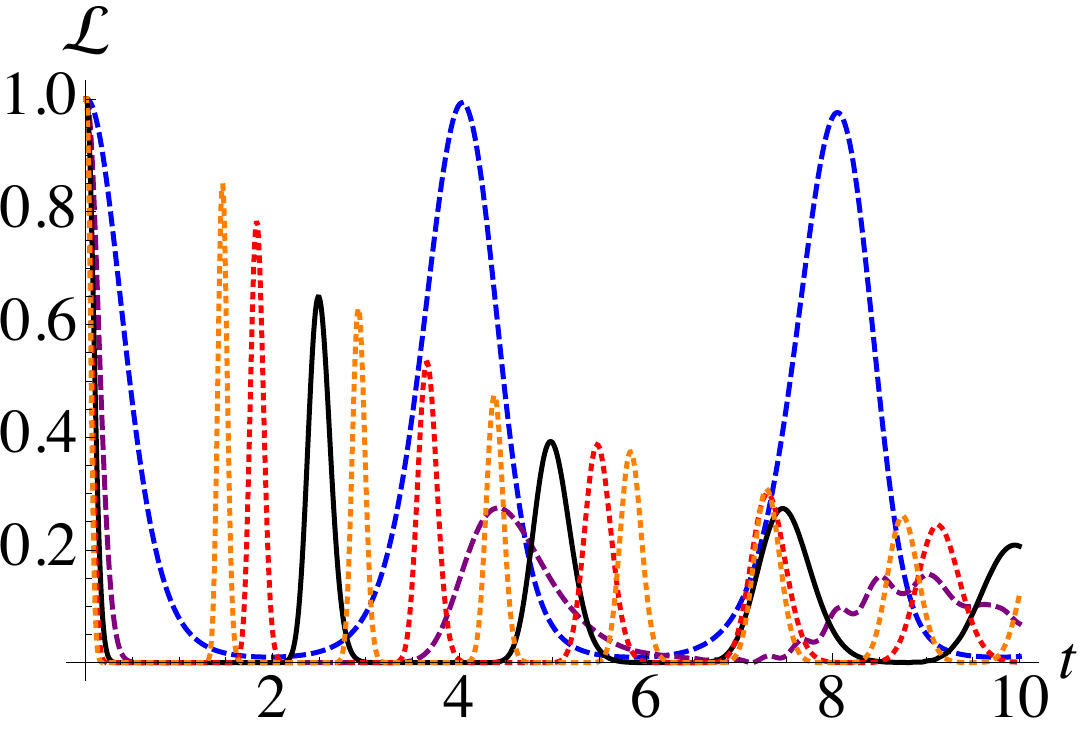}
\caption{(Color online) Time-dependent fidelity for a quench from $h_i=0.5$ to $h_f \in [0.6,1.4]$ with $N=400$. The dashed curves are for $h_f<1$ with $h_f=0.6$ (blue) and $h_f=0.8$ (purple). The dotted curves are for $h_f>1$ with $h_f=1.2$ (red) and $h_f=1.4$ (left-most, orange). The solid black curve is for $h_f=1$.}
\label{fig5}
\end{figure}

\subsection{From the Ferromagnetic to the Paramagnetic Phase}
We next consider the complementary case of beginning in the ferromagnetic phase, setting $h_i=0.5$, and quenching to increasingly larger values of $h_f$. This scenario is markedly different to that of the previous section as the ground state is now {\it nearly} degenerate, i.e. there is a exponentially (in $N$) vanishing gap between the ground state the the first excited state, cfr. Fig.~\ref{fig1}. Therefore, by quenching the field strength and kicking the system out-of-equilibrium, it quickly becomes excited and occupies higher order states. In Fig.~\ref{fig5} we see this effect clearly, contrary to the previous section, we see even for moderately small quenches ($h_f\gtrsim0.6$) the TDF reaches zero, indicating the evolved state is orthogonal. We further remark, the larger the system the smaller the quench required to achieve dynamical orthogonality, again this is a consequence of the fact that the energy gap between the ground and first excited states decreases with increasing $N$. Thus we cannot use the presence of orthogonality to witness signatures of the QPT in the dynamics. However, there is a clear qualitative difference appearing when the quench is near to or above the critical point. For quenches to $h_f\gtrsim 1.0$ the TDF evolves into fully orthogonal states for a period, before exhibiting short time revivals. The height of these peaks are steadily decreasing in the considered time window, and the width of the revivals broadens. This indicates a sizeable increase in the irreversibility of the process when quenches into the paramagnetic phase are considered.

\begin{figure}[t]
{\bf (a)}\\
\includegraphics[width=0.8\columnwidth]{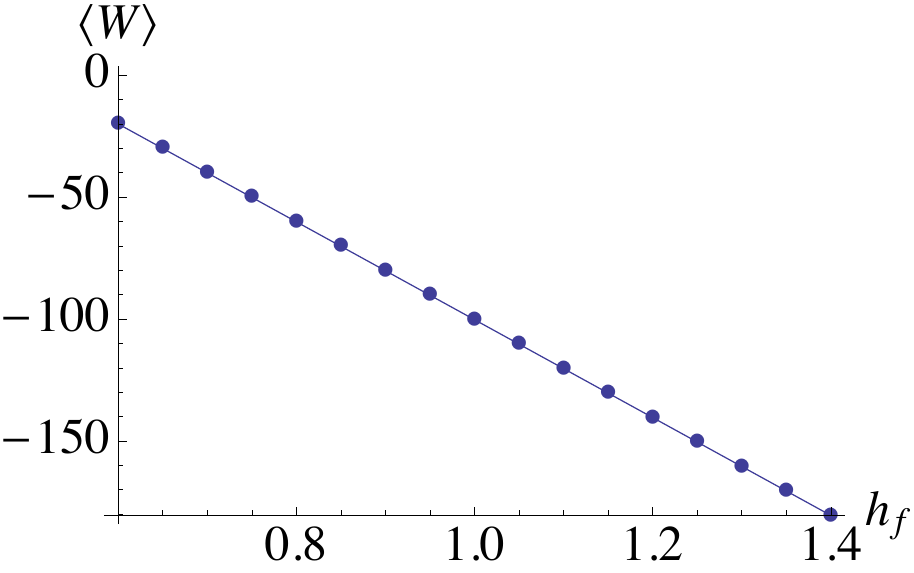}\\
{\bf (b)}\\
\includegraphics[width=0.8\columnwidth]{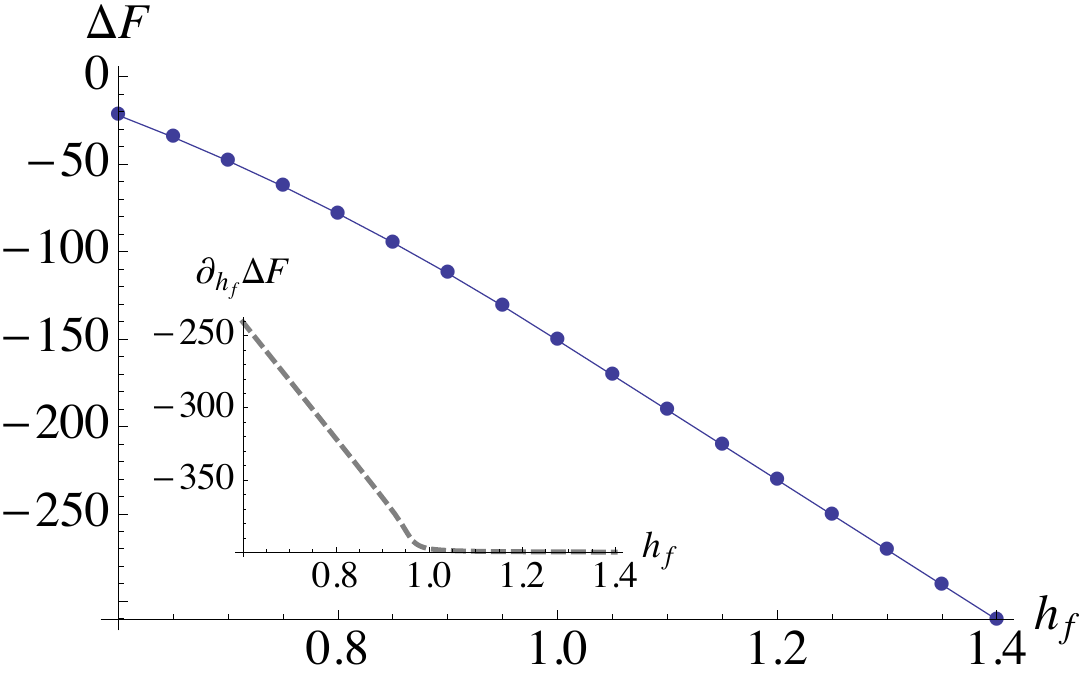}\\
{\bf (c)}\\
\includegraphics[width=0.8\columnwidth]{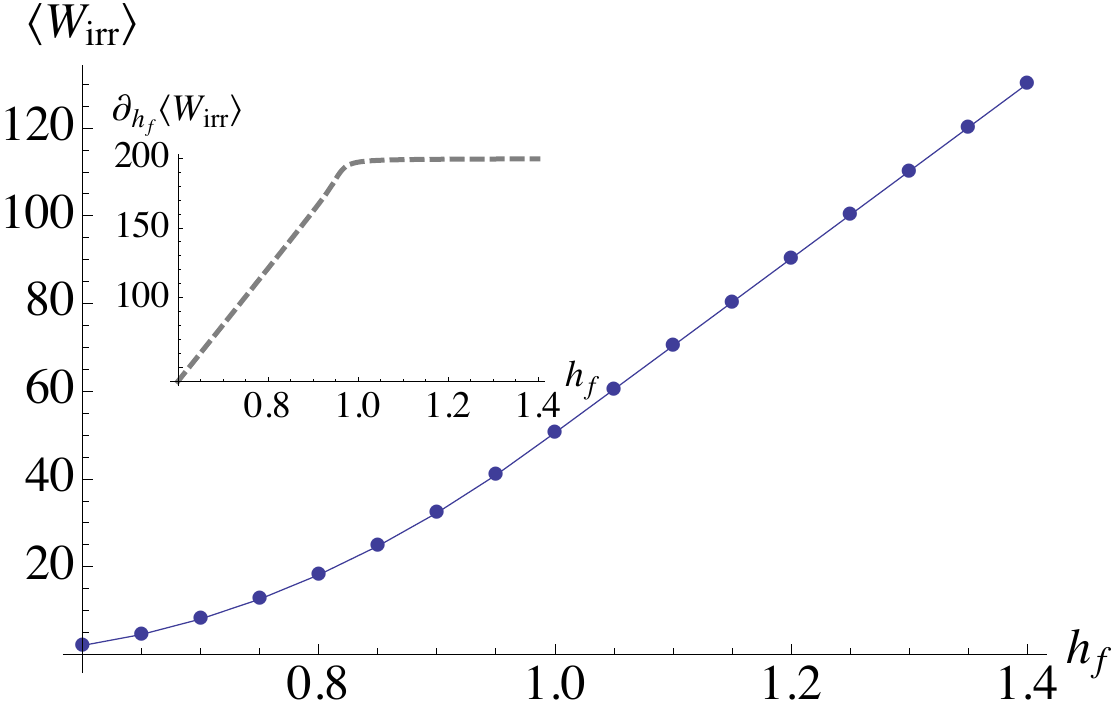}
\caption{(Color online) {\bf (a)} Average Work {\bf (b)} Free energy {\bf (c)} Irreversible work for a quench from $h_i=0.5$ to $h_f$. The insets of panels {\bf (b)} and {\bf (c)} are the first derivative of the functions. In all panels $N=400$.}
\label{fig6}
\end{figure}

We confirm this behavior in Fig.~\ref{fig6}. Panel {\bf (a)} shows that the work is a linear function of the magnitude of the quench. However, the free energy and the irreversibility show the same qualitative behavior as shown in the previous section. Focussing on the average irreversible work, due to the vanishingly small gap between the ground and first excited states, we see even small quenches are accompanied by a degree of irreversibility, the rate of which grows as the magnitude of the quench is increased. However, beyond the critical point we find the irreversibility grows linearly with the size of the quench.

\section{Analysis Based on the Spectral function}
\label{spectral}
\begin{figure*}[t]
{\bf (a)} \hskip1\columnwidth {\bf (b)}\\
\includegraphics[width=1.0\columnwidth]{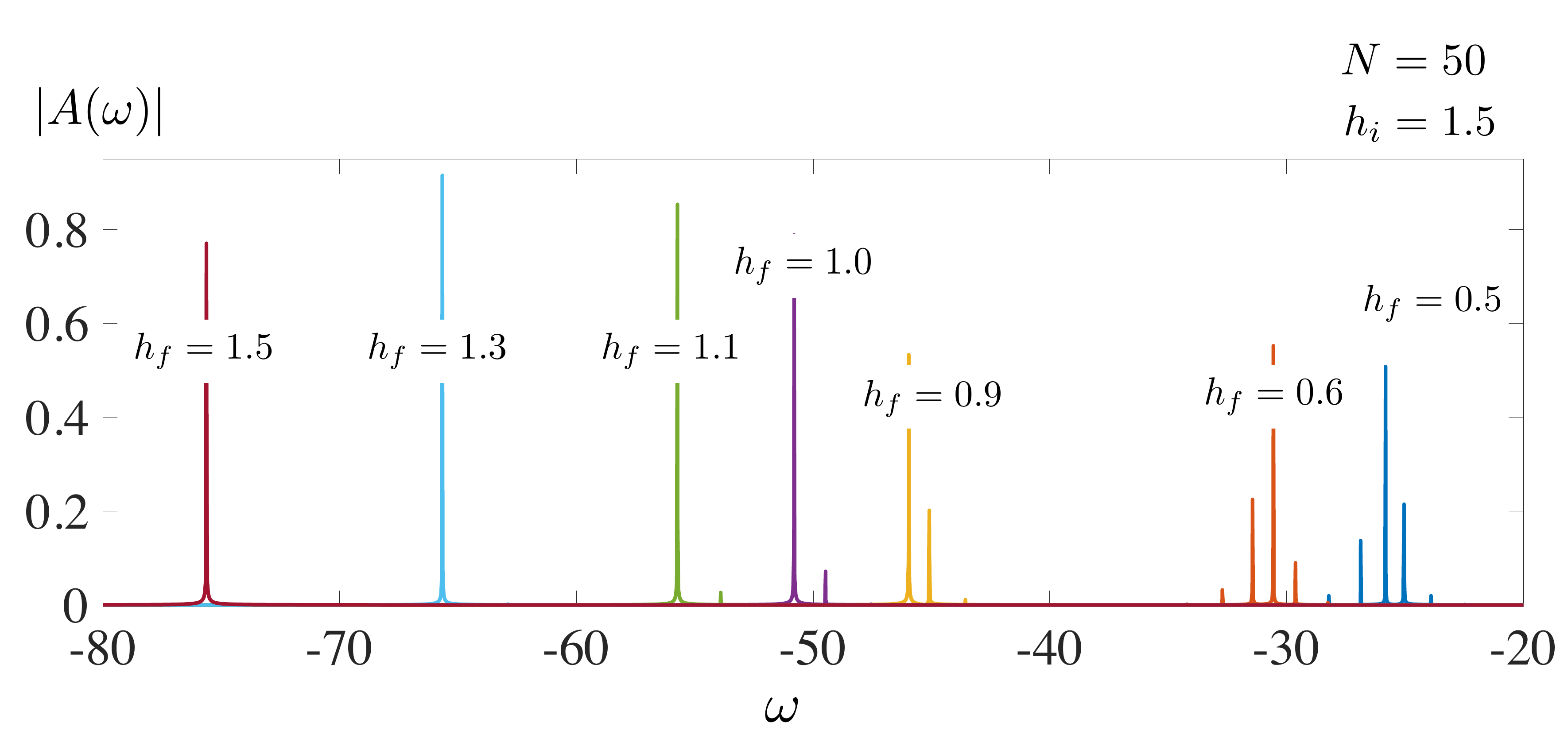}~
\includegraphics[width=1.0\columnwidth]{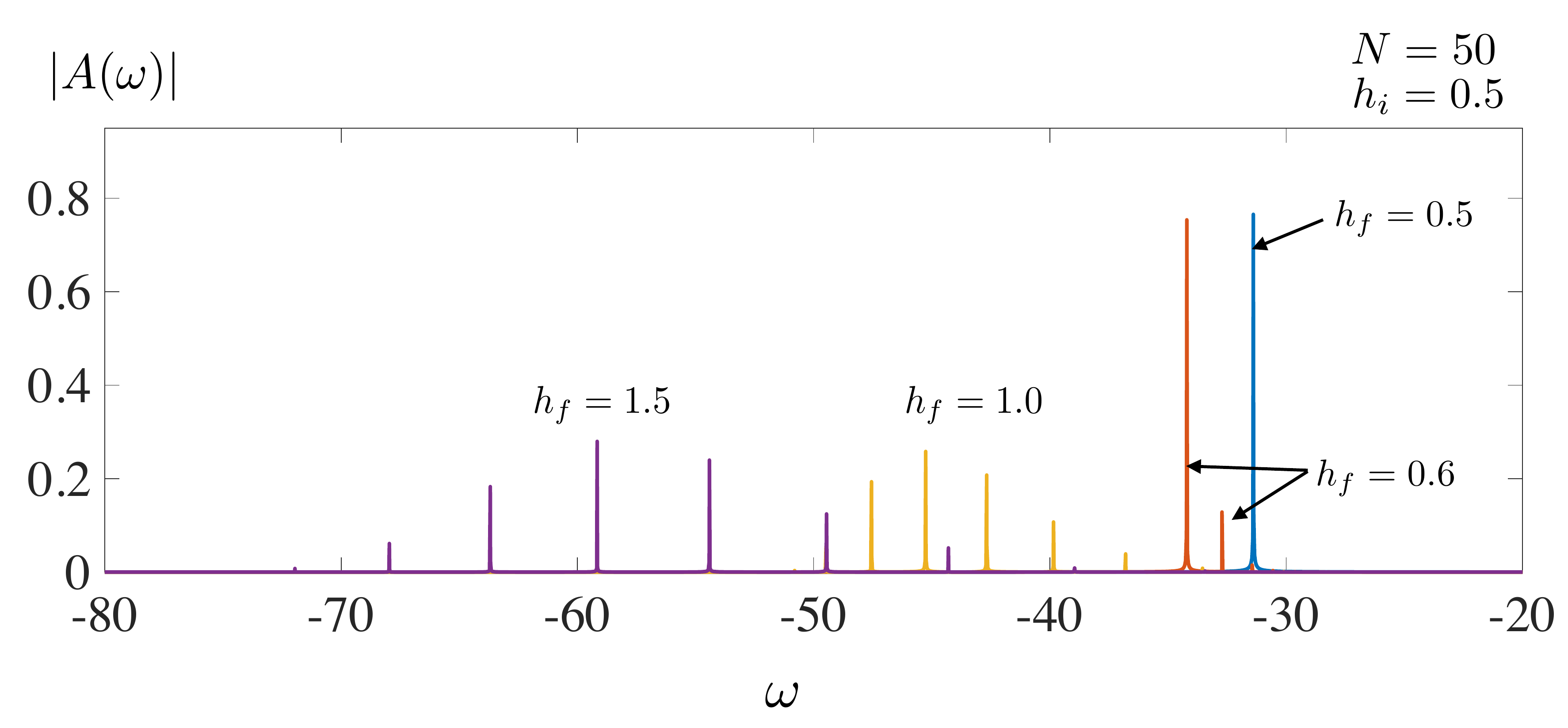}
\caption{(Color online) Spectral function, Eq.~\eqref{specEq}, for $N=50$. {\bf (a)} Starting in the paramagnetic phase with $h_i=1.5$ we examine the spectral function for several decreasing values of $h_f$. {\bf (b)} Spectral function for several quenches to $h_f$ starting from the ferromagnetic phase with $h_i=0.5$.}
\label{fig7}
\end{figure*}

Finally assess the behaviour of the spectral function (SF)
\begin{equation}
\label{specEq}
A(\omega) = 2 \Re \int e^{i \omega t} \mathcal{O} dt.
\end{equation}
This gives insight into the fundamental excitations that are governing the evolution, and therefore serves as an informative tool in understanding the the dynamics of the system~\cite{Mossy}. In Fig.~\ref{fig7} we show the SFs for several quenches, both when the quench remains in the same initial phase, and when it is across the critical point. We restrict ourselves to $N=50$ for simplicity, although qualitatively similar results hold for larger systems. 

In panel {\bf (a)} we assume the system begins in the paramagnetic phase fixing $h_i=1.5$. For no quench, i.e. $h_f=1.5$, the SF is a single peak exactly at the ground state energy. For small quenches staying within the paramagnetic phase we see the dynamics continues to be dictated only by the ground state, and this helps understand why the process is fully reversible. As we approach the critical point a second peak appears, corresponding to the second excited state of the final Hamiltonian. When $h_f \sim 1$ this second peak becomes more prominent. Quenching into the ferromagnetic phase, we see significantly more levels enter into the dynamics of the system. Interestingly although the the ground state still contributes to the dynamics, higher excited states play a significantly more dominant role  and this results in dynamical orthogonality. We remark, the model naturally has two distinct subspaces. Since our initial state is in the even excitation subspace, only the even states play a role in the dynamics.  

In Fig.~\ref{fig7} {\bf (b)} we show the complementary analysis starting from the ferromagnetic phase $h_i=0.5$ and quenching to larger values of the field. Again for reference, we see when no quench is performed, the SF is a single peak exactly at the ground state energy. However, now even for small quenches, $h_f=0.6$, due to the significantly more dense energy spectrum in this phase, more (even excitation) levels play a role in dictating the dynamics of the system, and therefore the system almost immediately witnesses dynamic orthogonality. This in turns allows us to understand the significantly larger irreversibility of the process when quenching from the ferromagnetic into the paramagnetic phase. However, similar to the previous case, when the quench remains in the same phase, i.e. $h_f \leq 1$ in this case, the ground state is still dominant. Quenching (near) to the critical point we see the SF spreads. For $h_f$ deep in the paramagnetic phase the SF is very spread, and again since we have quenched through the QPT, we see higher excited states play the most dominant role in the dynamics. 

\section{Discussions and Conclusions}
\label{conclusions}
We have examined the dynamics arising by quenching the parameters of the many-body interacting Lipkin-Meshkov-Glick (LMG) model. Starting from the ground state in a particular phase, and using the time dependent fidelity, we have shown that manifestly different dynamics occur when the quench is restricted to the same initial phase compared to a quench through the critical point. By employing tools from quantum thermodynamics we have shown that the average work maintains a linear relationship with the magnitude of the quench, regardless if it is through the quantum phase transition (QPT) or not. In contrast, the free energy and irreversible work are acutely sensitive to this difference. We find that quenching through the QPT leads to significant increases in the degree of irreversibility. This result can also help in understanding why controlling such many-body systems is so difficult through their QPTs~\cite{AdolfoPRL,CampbellPRL,SaberiPRA}, as it is this irreversibility that needs to be controlled. Starting from the paramagnetic phase, where there is a sizeable energy gap between the ground and first excited states, we have shown the occurrence of dynamical orthogonality serves as a remarkable witness of criticality in the model. Furthermore, by examining the spectral function we have shown that when the quench is through the critical point, the fundamental excitations that govern the dynamics are no longer dictated primarily by the ground state, but in fact higher excited states play the most prominent role. It is important to remark that our analysis is restricted to zero temperature and a natural question arises regarding the situation for finite temperatures. In the case of a quench from the paramagnetic to the ferromagnetic phase, the presence of the energy gap means that for reasonably small temperatures (i.e. temperatures that fail to provide enough thermal energy to excite the first excited state) the results remain largely unaffected. Conversely, when quenching from the ferromagnetic phase, due to the vanishingly energy gap even small temperatures lead to the first excited state becoming populated and thus can significantly change the dynamics. Our results highlight the interesting role the static properties of a many-body system can play in its dynamics. Indeed, such a role has recently been explored in Refs.~\cite{Heyl1,Heyl2,Heyl3,Heyl4,Heyl5} where, for the Ising model and also long-range interacting spin models (including the LMG model), so-called ``{\it dynamical quantum phase transitions}'' have been characterized. Our results add further evidence that equilibrium QPTs are clearly manifest in non-equilibrium processes. Finally we remark that we expect similar features to appear for other 1-dimensional spin systems such as the Ising model, however we leave this for a future study.

\section*{Appendix - Finite Anisotropy}
\begin{figure}[h!]
{\bf (a)}\hskip0.35\columnwidth{\bf (b)}\\
\includegraphics[width=0.49\columnwidth]{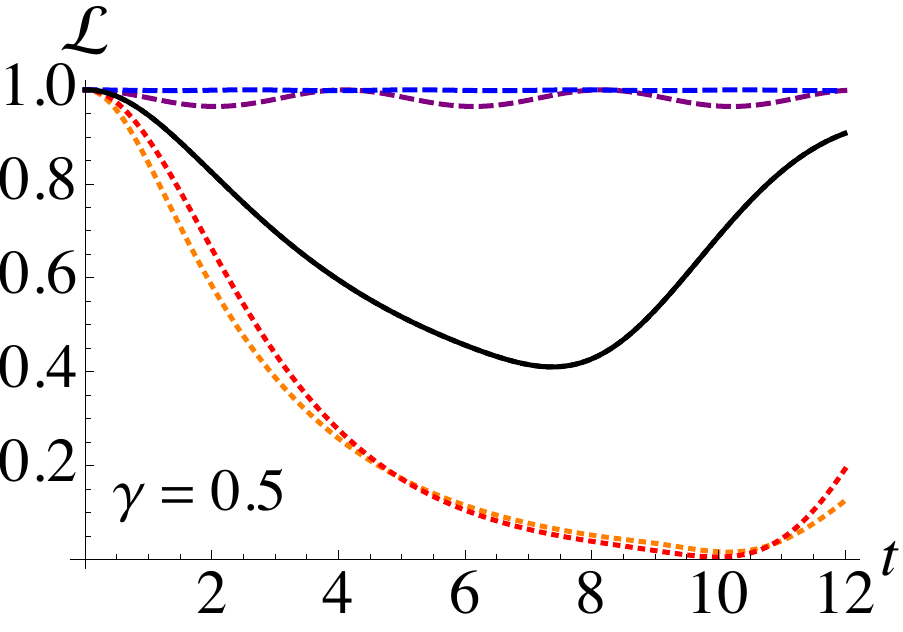}
\includegraphics[width=0.49\columnwidth]{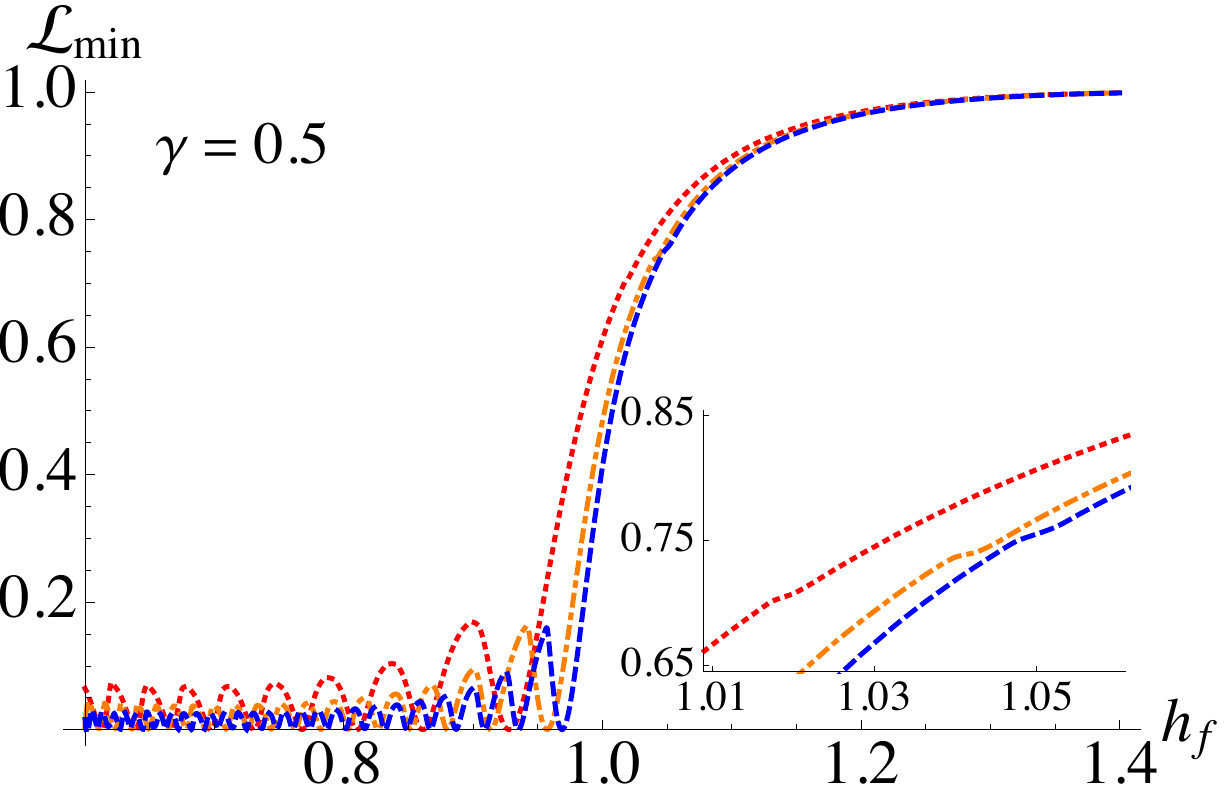}
\caption{(Color online) Finite anisotropy, $\gamma=0.5$: {\bf (a)} Time-dependent fidelity for a quench from $h_i=1.5$ to $h_f \in [0.6,1.4]$ with $N=300$. The dashed curves are for $h_f>1$ with $h_f=1.4$ (topmost, blue) and $h_f=1.2$ (purple). The lowest two dotted curves are for $h_f<1$ with $h_f=0.8$ (red) and $h_f=0.6$ (bottom-most orange). The solid black curve is for $h_f=1$. {\bf (b)} {\it Main Panel:} minimum value of TDF achieved for $t\in(0,12)$ quenching the field from $h_i=1.5$ to $h_f$. Each curve from left to right corresponds to an increasing size of $N=100$ (dotted, red), 200 (dash-dotted, orange), and 300 (dashed, blue). {\it Inset:} Zoomed in cross-section of main panel.}
\label{fig_appendix}
\end{figure}

Here we examine a finite value for the anisotropy parameter $\gamma$ showing that qualitatively the results in the main text are unaffected. We choose $\gamma=0.5$ and restrict to the case of quenching from the paramagnetic to the ferromagnetic phase for brevity. In Fig.~\ref{fig_appendix} we show the TDF [panel {\bf (a)}] and its corresponding dynamical minimum [panel {\bf (b)}], which are complementary to the results shown in Figs.~\ref{fig2} and \ref{fig3} of the main text. For finite $\gamma$ the TDF exhibits the same change in behavior when the quench is through the critical point. However, by changing $\gamma$ we are altering the energy of the system, and therefore this will be evidenced by a change in the frequency of the TDF. We clearly see this effect in panel {\bf (a)} as the time at which the first minimum is achieved is larger than in the $\gamma=0$ case. When the quench is restricted to a single phase the clean periodic behavior is maintained, while for values of $h<1$ this feature is lost and we find that the system can become dynamically orthogonal. Interestingly the `kink' in $\mathcal{L}_\text{min}$ is also still present.

\acknowledgements
I am grateful to Gabriele De Chiara, Thom\'as Fogarty and Mauro Paternostro for useful discussions and exchanges. This work is supported by  the EU Collaborative Project TherMiQ (Grant Agreement 618074), the Julian Schwinger Foundation (JSF-14-7-0000), and COST Action MP1209 ``Thermodynamics in the quantum regime".

\end{document}